\documentclass[preprint,review,11pt]{elsarticle}

\usepackage{graphicx}
\usepackage{amssymb}
\usepackage{subfigure}
\usepackage{amsmath}
\usepackage{amssymb}
\usepackage{listings}
\usepackage{graphicx}
\usepackage[rightcaption]{sidecap}
\usepackage{textcomp}
\usepackage{array} 
\usepackage{booktabs} 
\usepackage{multirow} 
\usepackage{caption} 
\usepackage{floatflt}
\usepackage{subfigure}
\usepackage{rotating} 
\usepackage{scrextend}
\usepackage[ruled,vlined]{algorithm2e}
\usepackage{listings}
\usepackage{xcolor}
\usepackage{url}

\usepackage[figurename=Fig.]{caption}
\usepackage{lineno}

\biboptions{numbers,semicolon}
\journal{Journal of The Royal Society Interface}

\begin{document}

\begin{frontmatter}


\title{Bottom-up robust modeling for the foraging behavior of \textit{Physarum polycephalum}}


\author[label1]{Damiano Reginato}
\ead{damiano.reginato@gmail.com}

\author[label1]{Daniele Proverbio}
\ead{daniele.proverbio@unitn.it}

\author[label1]{Giulia Giordano\corref{cor1}}
\ead{giulia.giordano@unitn.it}

\address[label1]{Department of Industrial Engineering, University of Trento, via Sommarive 9, 38123 Trento}

\cortext[cor1]{Correspondence: giulia.giordano@unitn.it}

\begin{abstract}
The true slime mold \textit{Physarum polycephalum} has the remarkable capability to perform self-organized activities such as network formation among food sources. Despite well reproducing the emergence of slime networks, existing models are limited in the investigation of the minimal mechanisms, at the microscopic scale, that ensure robust problem-solving capabilities at the macroscopic scale. To this end, we develop three progressively more complex multi-agent models to provide a flexible framework to understand the self-organized foraging and network formation behaviors of \textit{Physarum}. The hierarchy of models allows for a stepwise investigation of the minimal set of rules that allow bio-inspired computing agents to achieve the desired behaviors on nutrient-poor substrates. By introducing a quantitative measure of connectedness among food sources, we assess the sensitivity of the model to user-defined and bio-inspired parameters, as well as the robustness of the model to parameter heterogeneity across agents. We ultimately observe the robust emergence of pattern formation, in line with experimental evidence. Overall, our study sheds light onto the basic mechanisms of self-organization and paves the way towards the development of decentralized strategies for network formation in engineered systems, focusing on trade-offs between biological fidelity and computational efficiency.
\end{abstract}


\end{frontmatter}


\section{Introduction}

Self-organization is the spontaneous ability of organisms to perform global tasks through an ensemble of local actions and adaptation. A leading specimen for studies about emergent problem-solving ability and self-organization is the true slime mold \textit{Physarum polycephalum}, a member of the Myxogastria class of organisms, also known as acellular slime molds; in fact, in its plasmodium (vegetative) stage, \textit{P. polycephalum} is composed of a single cellular membrane enveloping a multitude of nuclei. In biology and complex systems science, \textit{Physarum} is extremely valuable to study cell cycle \citep{cellCycles} and DNA replication \citep{DNAreplication}, as well as motility \citep{motility}, environmental sensing \citep{hader1984phototactic}, and response to chemical and physical stimuli \citep{Ueda1975}. 

In its vegetative stage, the plasmodium moves through a peculiar process called shuttle-streaming \citep{oettmeier2017physarum}. The cytoplasm differentiates into two forms: rigid gel-like ectoplasm (gel, a stiff sponge-like matrix composed of actin–myosin fibers) and fluid endoplasm (sol) flowing within the matrix. Inside vein-like structures of ectoplasm, periodic oscillations of contractile fibrils cause back-and-forth movements of the endoplasm, making the plasmodium elongate and move. When a local concentration of nutrients is sensed, the external membrane softens in its proximity, making the streaming more effective in that direction and thus allowing the organism to reach the food source more quickly \citep{rodiek2015patterns}. Initially, a fine mesh of tubules (well approximated by a planar sheet) covers the area until the food source is engulfed by the plasmodium. Then, the network structure changes to efficiently distribute the nutrients throughout the organism. The more nutrient flows through a \textit{vein}, the thicker this becomes, in a reinforcing feedback loop that ultimately leads to a mature network configuration \citep{littleTubesGone}. 

Shuttle-streaming enables \textit{P. polycephalum} to showcase emergent computing ability: it can successfully link multiple food sources \citep{Nakagaki2004} while avoiding risky environments \citep{Nakagaki2007}, solve puzzles and mazes by connecting food sources via shortest paths \citep{Nakagaki2000}, and optimize \textit{in-vitro} replicas of transport networks \citep{Jones2011}. Given its unique ability to solve complex problems without possessing central brain-led coordination, \textit{P. polycephalum} inspires experiments and models about swarm intelligence, emergent computing and self-organization \citep{Beekman2015, Reid2023}. It also serves as bio-inspiration for the development of efficient algorithms for optimization, maze exploration and computing \citep{tsompanas2014evolving, Adamatzky2016}, as well as for artificial swarm systems \citep{jones2016applications, yang2019bio}.

To capture the mechanisms underpinning the emergence of complex problem-solving behaviors, numerous models of \textit{P. polycephalum} have been proposed, at various scales and with different scopes \citep{oettmeier2017physarum}. Fluid-dynamics models studied directed migration and memory brought about by oscillations and peristalsis within the cellular endoplasm \citep{Rodiek2015, Alim2013}; network formation was reproduced using e.g. Hagen-Poiseuille laws on graphs \citep{Tero2007} or reaction-diffusion models \citep{adamatzky2012slime, Ghanbari2023}; optimal path formation was demonstrated using memristor-like network dynamics \citep{Pershin2009,Gale2015}; and bio-inspired models of computing systems have been surveyed e.g. in \citep{liu2017new}. These models captured several aspects of \textit{Physarum} evolution, such as the influence of peristalsis and fluid dynamics in its growth, or the role of nonlinear threshold responses to stimuli in driving optimality in network formation. However, they lack a bridge through scales, to inquire in detail how microscopic interactions lead to the emergence of patterns at macroscopic scale -- and how disturbances are coped with, at any scale, leading to robust behaviors.

To study macroscopic behaviors emerging from microscopic interactions, multiscale models have been developed, like Cellular Automata \citep{Jones2011, Wu2015, gunji2008minimal}, having low complexity and thus low biological fidelity, or Multi-Agent Systems (MASs) \citep{liu2017new}, having higher complexity but potentially higher computational cost, to investigate point-of-interest and foraging behaviors of the whole organism, driven by local decision-making at the microscopic level.
As a fundamental advantage, MASs allow to add layers of complexity that go beyond simpler individual-based approaches. For instance, the decision-making rules of individual computing units can be made more realistic by incorporating knowledge of the biological and mechanical phenomena; multiple self-organizing behaviors can be integrated; and dynamics over multiple scales can be concurrently tracked \citep{proverbio2024chemotaxis}. Going beyond what has been discussed in previous works, MASs can also be fruitfully employed to study the robustness of emergent patterns against uncertainties and disturbances at any scale, and thus assess under which stress conditions a self-organizing entity manages to thrive.

However, designing a MAS to model the behavior of \textit{P. polycephalum} requires a careful interpretation of the modeled agents: no easy parallel can be drawn between a collection of independent agents, having a granular nature, and the whole \textit{Physarum} organism in its plasmodium phase. Nonetheless, following an approach discussed in \cite{Jones2011, liu2017new, wu2012enhanced}, agents can be thought of as units of solid cytoplasm being transported through the fluid endoplasm (gel/sol interaction).
Particle-based models with a bottom-up approach could help elucidate the basic computing mechanisms required to enable network formation and foraging; however, it is still unclear which are the minimal behavioral rules needed to reproduce foraging as a combination of two self-organized mechanisms -- search and networking -- along with efficient pruning and nutrient transport.
Moreover, recent models \citep{liu2017new} have considered diffusive food sources, to exploit \textit{Physarum}'s chemotactical capability of streaming up gradients, while nutrients used in experiments (such as oat flakes) or present in real life are typically little or not diffusive \citep{Jones2010}. Hence, the question remains of how \textit{P. polycephalum} manages to search a more realistic environment and complete its foraging task.

In this work, to investigate the minimal mechanisms leading to self-organized efficient network formation, we develop and test three competing agent-based models of increasing complexity and realism; we analyze their performance, their sensitivity to parameter values and their robustness to heterogeneity in the parameters. We focus on maze-free environments with discrete and non-diffusive food sources, and we observe the formation of connections between the food sources, akin to those shown in the experimental photo in \cite[Fig. 1]{Jones2010}. We also introduce quantitative metrics to automatically check the formation of robust connections in our simulations, and to quantify the network efficiency (i.e., its closeness to a direct, straight connection between the food sources). Our results shed light onto emergent mechanisms for spontaneous foraging and pattern formation, and pave the way towards the development of bio-inspired algorithms.

\section{Methods}

\textit{Physarum polycephalum} gives rise to self-organized network formation when foraging (i.e., searching
for nutrients in a large area) and transporting nutrients efficiently, by identifying short transport distances and maintaining fault tolerance against disconnections. We adopt a multi-agent approach to reproduce the spontaneous emergence of such behaviors, as a combination of self-organized \textit{exploration} and \textit{networking} dynamics. We embed active agents \citep{jones2009passive}, able to modify their local surroundings by releasing chemoattractants, in a dynamic environment with discrete and non-diffusive food sources, and we perform in-silico experiments using the GAMA Platform IDE \citep{grignard2013gama}. Suitable metrics to assess the quality of connections and network formation are also introduced to perform sensitivity and robustness analysis.

\subsection{Three competing models}

In line with the MAS approaches described above, we interpret the model agents as units of interaction between solid cytoplasm and fluid endoplasm that flow inside the tubules. In a computational framework, we thus interpret the \textit{Physarum} organism as a collection of single computing elements (described in detail below), each able to perform tasks based on individual behavioral rules, and we look for the minimal set of rules yielding the desired emergent behaviors. Multi-Agent Systems offer the natural framework for this endeavor, allowing us to design and simulate models composed of several computing elements, and also to test their robustness, meant both as the model's ability to preserve its qualitative behavior in spite of uncertainties \citep{blanchini2021structural} and as the computing system's ability to maintain its  function despite alterations of individual programs.

A MAS model is a tuple $\mathcal{M} = \langle \{A_k\}_{k=1}^N, \{P_k\}_{k=1}^N, E \rangle$ of $N$ agents $A_k$, each following a set of behavioral rules contained in program $P_k$, and their environment $E$.
Here, we develop and test three different models that share the same environment settings, so that $E$ is fixed. Each model is characterized by a different program $P_i$ that is common to all the agents: $\mathcal{M}_i = \langle \{A_k\}_{k=1}^N, P_i, E \rangle$, $i=1,2,3$. Therefore, with a slight abuse of notation, we use the terms \textit{model} and \textit{program} interchangeably. We analyze and compare the models, or programs, $\{P_i\}_{i=1}^3$, each composed of a set of rules that are derived from existing models of \textit{P. polycephalum} or inspired by its biological functions. The three models are characterized by increasing levels of complexity, measured in terms of the number of rules followed by the agents \citep{zenil2019compression}. In particular:

\begin{enumerate}
\item The \textit{networking model} (NM) only reproduces the networking phase of foraging, while it lacks the exploration phase. The agents, initially scattered randomly in the entire environment, aggregate to form a network. We use NM as a benchmark, since it relates to important models in the literature \citep{Jones2010, Jones2011, Wu2015}. 
\item The \textit{exploration model} (EM) includes both phases of the self-organized foraging process: the agents, starting from a predetermined area within the environment, first explore and locate food sources, and then aggregate to form a network. Inspired by \citep{liu2017new}, we consider two possible states for the agents, $\sigma_1$ (exploration) and $\sigma_2$ (networking), associated with different behavioral rules. The two states describe the evolution of the plasmodium, which first elongates under peristaltic movement, and then forms a network by thickening the most trafficked veins and letting the others die out. The switch between the two states is governed by a local \textit{starvation} condition: an agent that cannot find food for a given number of cycles will stop exploring the environment and instead actively search for higher chemoattractant concentrations. This condition expresses the organism's need to maximise the explored area while balancing internal energy during shuttle-streaming, and differentiates EM from the models proposed in the literature.
\item The \textit{life-cycle model} (LCM) implements a full life cycle for the gel/sol units (agents), which can dynamically be generated and die; the number of agents, initially small, thus varies over time. The agents are initially in state $\sigma_1$ (exploration) and can switch to state $\sigma_2$ (active food location and networking) following the \textit{starvation} condition. Food sources are hubs for reproduction: if the density of agents finding food within their grid cell is sufficiently low, agents duplicate (each generates a copy of itself). Hence, the population grows and connections can be established more rapidly, as the feedback loop between nutrient relay and flow through a vein is strengthened. If starving agents do not find food within a given period of time, they die: only branches that connect food sources survive, while others wither and disappear. Agents can also die when the local density of neighboring agents is too high (and hence they cannot move and survive). This mechanism, inspired by \cite{liu2017new}, promotes local autonomy and overall self-organization. LCM is the most complex model, and the closest to the biological system, in that it also considers demographic dynamics.
\end{enumerate}

We now provide further implementation details for each of the three models.

\subsection{Environment}

For all the models, we consider the same \textit{planar} environment, which is justified because the plasmodium is very thin and experiments are mostly conducted in a 2D setting \citep{adamatzky2012slime}. In particular, our square environment of side $d \in \mathbb{N}$ is subdivided into a virtual $d \times d$ grid of cells. Each cell can host more than one agent at a time (in light of the fluidic interpretation of agents). The cell size is scaled to approximate typical dimensions used in experiments: drawing from experimental figures in \citep{Nakagaki2004,900}, we set $d = 80$ unit cells and a cell size of 0.1875 mm, so that the square side is 15 mm long. Each cell $(i, j)$ is associated with two scalar values, the amount of \textit{food} $F_{i,j}$ and the amount of \textit{trail tokens} $T_{i,j}$ (chemoattractants secreted by the agents). The agents are constrained to move within a \textit{circular} subset of the environment, which is the maximum-area circle inscribed in the square, to mimic a common choice of substrate shape in experimental settings \citep{Nakagaki2004, stepwise}, and boundary conditions are fixed. Two non-diffusive food sources are located at the top and the bottom of the environment; their size can be chosen arbitrarily without loss of generality (changing it would amount to rescaling the whole environment). Each food source represents a discrete piece of food (usually oat flakes, either whole \citep{adamatzky2010road} or powdered \citep{Nakagaki2004}, in experimental settings) placed upon a nutrient-poor substrate. Food sources induce prescribed levels of chemoattractants in the environment: a cell $(i,j)$ within radius $r_F$ from the food source contains an amount of food chemoattractant
\begin{equation}
    F_{i,j} = \frac{F_{\text{max}}}{1+(r_{i,j}^k)^2} \ ,
\end{equation}
where $F_{\text{max}}=15$ is the maximum food quantity and $r_{i,j}^k$ is the distance of cell $(i,j)$ from the center of the $k$-th food source. To mimic a realistic pile of powdered flakes, we set $r_F=7$
units, corresponding to 1.3 mm.
The level of trail chemoattractant $T_{i,j}$ in each cell depends on the agents' presence and behavior (see Section \ref{sec:agents}).
Given the value of $F_{i,j}$ and $T_{i,j}$ for all the cells, the agents move from one cell to another in the environment either by random exploration, or by following chemoattractants (both food and trail tokens), thus yielding emergent interconnections among agents and between agents and food sources.

\subsection{Agents}
\label{sec:agents}
Following \cite{Jones2011}, we model the plasmodium's multinucleate syncytium as a collection of identical components, representing a hypothetical unit of gel/sol interaction. \textit{Physarum} moves driven by hydrostatic pressure and a rhythmic streaming flow of sol throughout the matrix, which causes changes in plasmodium thickness. Pressure and shuttle-streaming are, in turn, governed by the resistance provided by the gel matrix to the flow of sol, and the spontaneous contraction of the actin–myosin fibers. This combination of mechanisms ultimately yields distributed oscillations. In the agent-based model, the movement of the particles replicates the production and flow of sol, while the resistance of the gel matrix is provided by overlapping sensory couplings and collisions between agents, which make the flow in ``crowded'' veins more hectic.  Moreover, sensing of food sources and chemotaxis (movement along chemical gradients) are governed by threshold phenomena, as the slime mold responds to environmental stimuli with step-wise changes in plasmodium shapes \citep{Ueda1975, Ghanbari2023}. Agents thus possess minimal mechanisms for sensing and automatic orientation, in response to \textit{if-then} stimuli. The final structure of the \textit{Physarum} transport network is given by the collective position of the agent population, whose movement represents the flow of protoplasmic sol within the organism.

\begin{figure}[h!]
\centering
\includegraphics[width=\linewidth]{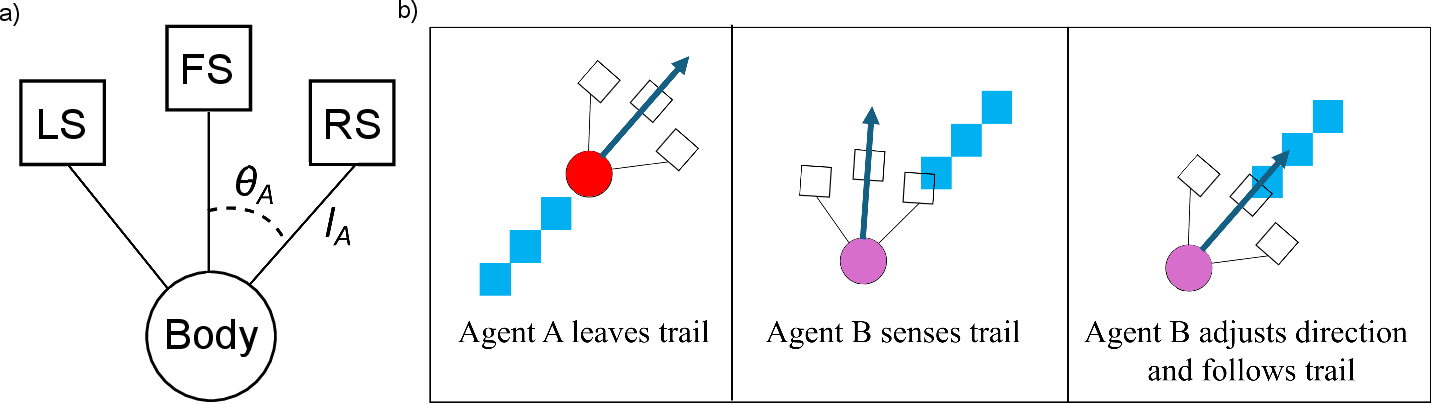}
\caption{\small a) Agent morphology. Main body with three sensors: left, forward and right. The sensor arm length is $l_A$, while $\theta_A$ is the sensor arm angle; the positions of the left and the right sensors are symmetric with respect to the forward sensor. Each agent is much smaller than a cell, to allow for the overlapping of agents and chemicals, as described in the main text. b) Schematic visualization of interactions among agents, mediated by secreting and sensing chemical trails; agent B follows the logic of an agent in state $\sigma_2$. Agents are in red/purple, and the trail is in light blue.}
\label{fig:agent_struc}
\end{figure}

Similarly to \cite{Jones2011, liu2017new}, we model agents as virtual entities with a body and three sensors (left, right and forward) that detect chemical traces (when overlapping with them) and guide chemotaxis; see Fig. \ref{fig:agent_struc}. Crucial agent parameters are the sensor arm length $l_A$, representing the “reach” of cytoplasm units to influences coming from the surrounding environment and the other units, and the angle $\theta_A$ between the forward sensor arm and the left/right sensor arms; their default values are set to $l_A = 2$ and $\theta_A = 45^\circ$ according to previous literature \citep{liu2017new}, and we will perform sensitivity analysis on both parameters.

Agents actively influence the surrounding environment by secreting non-diffusive \textit{trail tokens}, representing Ca$^{2+}$ ions that move through the tubular veins to simulate peristaltic oscillations. Other agents are attracted by trail tokens, which reproduces the phenomenon of autocrine chemotaxis: agents both secrete and sense the same chemoattractant. Trail tokens, whose total amount in each environment cell $(i,j)$ is $T_{i,j}$, are secreted in quantity $Q_T = 1$ at each time step during the networking phase, and decay with a constant evaporation rate $\mu_T = 0.7$ (default value), representing the physiological dilution of chemicals. The evaporation rate is assumed to be constant as a first approximation; physiology-dependent dilution may be plausible, but hard to embed into a functional form, and thus its investigation is left to future studies. Since the concentration of ions cannot grow indefinitely, we set a maximum amount of trail tokens in a single cell, $T_{\text{max}} = 3$. Parameter values are set consistently with similar models \citep{liu2017new}, and the sensitivity of simulation outcomes to these values will be investigated.

Basic NM agents are tuples $\mathrm{A} = \langle \text{\textit{Process}, \textit{Inspect}, \textit{Act}} \rangle$. \textit{Process} is a set of inner features \{\textit{States,Update}\}, where \textit{States} = \{$\sigma_2$\}, since the networking state (finding nutrients and performing chemotaxis) is the only possible, and \textit{Update} = $\{\mathrm{U}_{l} \colon \mathrm{Per} \mapsto \mathrm{S}_{j}\}$ includes rules to update the agent state according to perceptions (and is empty by default for NM agents). \textit{Inspect} = $\{\mathrm{I}_{i} \colon \mathrm{E}_{g} \mapsto \mathrm{Per}\}$ includes rules mapping the environment configuration to internal perceptions, i.e., sensing of chemicals (trail tokens and food) underneath the sensors. \textit{Act} = $\{\alpha_{h} \colon \mathrm{S}_{j} \mapsto \mathrm{E}_{g'}\}$ is a set of actions that depend on the agent state and affect the environment configuration; \textit{Act}$(\sigma_2)$ makes the agent (i) follow the direction of higher chemical concentration, by rotating the body in the direction of the corresponding sensor, (ii) avoid collisions by turning $90^\circ$ each time a sensor overlaps with the environment edge, (iii) secrete trail tokens by augmenting $T_{i,j} \to T_{i,j} + Q_T$ (and, if $T_{i,j} + Q_T > T_{\text{max}}$, then $T_{i,j} \to T_{\text{max}}$).

Building upon NM agents, EM agents are equipped with a second possible state $\sigma_1$ (exploration), so that \textit{States} = $\{\sigma_1,\sigma_2\}$. \textit{Act}$(\sigma_1)$ makes the agent (i) follow the direction of lower chemical concentration, or (ii) follow the current heading, if chemical concentrations across multiple sensors are equal, with a random perturbation within a cone of width $\pm \pi/8$, representing small deviations from straight paths observed in biological organisms. \textit{Update} includes a counter condition for starvation, which increases by 1 each time no food is senzed. When a starving threshold $T_S = 100 + \mathcal{N}(0, 10)$ is reached, the state switches from $\sigma_1$ to $\sigma_2$; the random value $\mathcal{N}(0, 10)$ introduces heterogeneity across the agents.

LCM further extends EM by introducing demographic dynamics, so that \textit{States} = $\{\sigma_1,\sigma_2,\emptyset\}$, where $\emptyset$ means agent's ``death'' and elimination. \textit{Update} includes a second threshold $T_D = 250$: if the starvation counter crosses $T_D$, \textit{Update} triggers the switch $\sigma_2 \to \emptyset$, thus the agent dies. A density check is also implemented: \textit{Inspect} includes a function to sense the number of neighboring agents $N_A$, and \textit{Updates} triggers the switch $\sigma_2 \to \emptyset$ whenever $N_A>T_N$, where $T_N$ is a density threshold. Defining such threshold allows to computationally implement the observation that, in a too crowded environment, agents may be eliminated to promote local autonomy; in the absence of empirical validation, we employ the same parameter value as in \cite{liu2017new}. As for agents' birth, \textit{Act} includes a function to generate another agent whenever the original agent is on a food source and there is space, in the immediate vicinity, for the offspring: $N_A < T_B$, where $T_B$ is a birth threshold. $T_B$ and $T_N$ are set following geometric considerations. Agents are virtual entities, but can use sensors to check whether other agents are present in an environment cell. The IDE implements Moore-type neighborhood \citep{grignard2013gama}, so offspring can exist if $T_B < 8$. Conversely, to set $T_N$, we consider the area around an agent to be overpopulated if the whole circle of radius $l_A$ is full; for default parameters, this corresponds to $T_N = 18$.

\subsection{Connection metrics}
In addition to qualitatively comparing emergent patterns of the simulated models to experimental figures, we introduce an automated quantitative method to assess the existence of connections among food sources and to measure their efficiency. We first associate the environment with a matrix $M \in \mathbb{R}^{n \times n}$, corresponding to the partition of the environment into a grid of $n \times n$ squares, as shown in Fig. \ref{fig:env_setting}. Entries $M_{ij}$ can have value 1 (when the corresponding square contains some agents) or 0 (when the corresponding square contains no agents). This maps the agent population onto a graph of which $M$ is the incidence matrix, where non-zero elements $M_{ij}$ represent links among agent-populated nodes. To set the size $n$ of $M$, we note that agents follow nearby agents if they step upon their trail, which should have sufficiently high level. Therefore, for a connection to be formed, the distance between leading and trailing agents should be lower than the space that trailing agents can cover before the trail decays. We wish to prevent the method from returning that the two food sources are disconnected while agents are still forming streams (although less dense) and we thus set $n$ accordingly. In view of the exponential decay and the constant evaporation rate $\mu_T$, the trail becomes negligible (below $10^{-6}$ times the original level $Q_T$) after $t_T = \ln(10^6) \cdot \mu_T$ cycles. In that time span, an agent traveling straight would cover $s_A = v_A \cdot t_T$ units. To be conservative, we assume that the movement occurs along the diagonal of the matrix. Hence, the side of the square associated with $M_{ij}$ should be $l_m = s_A \cdot 2^{-1/2}$, and the matrix size is thus $n = d / l_m$. With our default parameters, it must be $n = 12$. To account for uncertainties, and to have an odd number of cells (which is useful to identify perfectly straight connections between food centers), we use $n_1 = 11$ and $n_2 = 13$.

After mapping the population to matrix $M_1 \in \mathbb{R}^{n_1 \times n_1}$ and matrix $M_2 \in \mathbb{R}^{n_2 \times n_2}$, we test connectedness by checking whether paths connecting the two food sources exist across agent-populated nodes, using the common recursion method for shortest paths. At the end of a batch of 100 simulations, the test for connectedness allows us to count the number $N_Y$ of simulations in the batch that formed a connection between the two food sources. Whenever a connection is formed, we measure its efficiency using $K$, the index of the furthest column from the central column $K_S$ that includes part of a connection between food sources.
Efficiency, defined as closeness to a perfectly straight connection along the central column, can be quantified  as $\mathcal{E'}=|K - K_S|$.
To enable direct comparison of quantitative data obtained using two different mapping matrices, $M_1$ and $M_2$, we normalize $\mathcal{E'}$ by its maximum possible value, which depends on the matrix size ($\mathcal{E}^{\text{max}}_{1} = 6$, $\mathcal{E}^{\text{max}}_{2} = 7$), and we thus obtain the metric:
\begin{equation}
    \mathcal{E}=\frac{|K - K_S|}{\mathcal{E}^{\text{max}}_{1,2}}.
    \label{eq:efficiency_norm}
\end{equation}
Hence, $\mathcal{E}$ close to 1 refers to connections being curvy and possibly spanning the whole environment, while $\mathcal{E} = 0$ corresponds to a straight line connecting the food sources.

\subsection{Simulation protocol}

For all three models, simulations are initialized with default parameter values (Table \ref{tab:default}) and run for a fixed simulation horizon $T = 3000$, within which we observe that mature network configurations are formed and persist. NM is initialized with agents randomly scattered over the whole environment, and in the networking state; EM and LCM are initialized with agents located within a small area, about halfway between the food sources, and in the exploration state. The number of EM agents is the same as for NM, to enable direct comparison; the choice of the initial arrangement only impacts the early phase of the transient (see Fig. \ref{fig:sims2}). 

To conduct our sensitivity and robustness analysis, we vary one parameter at a time, while the others keep their default values. Sensitivity refers to deterministic homogeneous changes of parameter values, which are modified in the same way for all the agents; tested intervals are in Table \ref{tab:default}. Robustness involves defining uncertainty intervals $\Delta p_i$ for each parameter $p_i$ considered for the analysis; for each agent within a simulation, the value for $p_i$ is randomly drawn, independently, from a uniform distribution $\mathcal{U}(\hat{p}_i - \Delta p_i, \hat{p}_i + \Delta p_i)$, where $\hat{p}_i$ is the default value. The probabilistic assignment of values drawn from uncertainty intervals expresses the heterogeneity and parameter variability among biological entities. In particular, we test the system robustness to uncertainties in agents' speed, as gel/sol units may flow heterogeneously depending on vein thickness, transported nutrients and other sources of uncertainty. Moreover, heterogeneity in speed over transport networks is a known issue \citep{chen2011transport}. We consider a range of uncertainty amplitudes $\Delta v_A = [0, 0.05, 0.1, \dots, 0.7]$ around the default speed $\hat{v}_A = 0.7$.

\begin{table}[ht]
		\centering
        \small
		\caption{Parameter values used for the analysis, corresponding scaled values in experimental units, and intervals for sensitivity analysis. If not specified otherwise, values are in arbitrary units and 1 simulation step corresponds to 1 second; $d$ is included to compare the scales.\\ $^{(*)}$ For $v_A$, we do not perform sensitivity analysis within a deterministic interval of values, but robustness analysis for a range of uncertainty intervals $\Delta v_A$, as discussed in the main text.}.
		\label{tab:default}
		\begin{tabular}{cccc}
			\toprule 
            \textbf{Parameter} & \textbf{Value} & \textbf{Scaled Value} & \textbf{Interval}  \\
            \midrule
            $d$ & 80 & 15 mm & -\\
            \midrule
			$l_A$  &  2  & 0.375 mm & [1; 3] \\ 
			\midrule
			$\theta_A$ & 45$^\circ$ & 45$^\circ$ & [30$^\circ$; 60$^\circ$]  \\
            \midrule
            $\hat{v}_A$ $^{(*)}$ & 0.7 &  0.13 mm/s & $\Delta v_A \in [0 ; 0.7]$ \\
            \midrule$
            \mu_T$ & 0.7 s$^{-1}$ & 0.7 s$^{-1}$ & [0.5; 0.9] \\
			\bottomrule
		\end{tabular}
	\end{table}
 
\begin{figure}[h!]
\centering
\includegraphics[width=\linewidth]{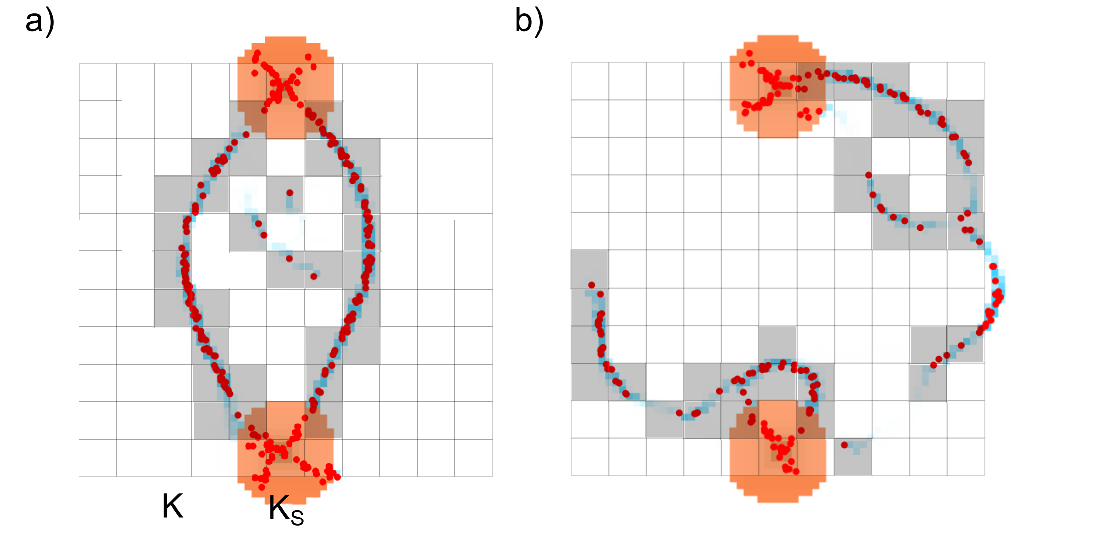}
\caption{\small Example of experimental setting, mapping to $M_1$, and connectivity check. Orange spots represent food piles, red dots represent LCM agents close to the end of a simulation. The $11 \times 11$ grid associated with matrix $M_1$ is superimposed. Shaded gray cells are those containing some agent, and thus associated with 1 entries in $M_1$; white cells correspond to 0 entries in $M_1$. a) The connectivity test is positive: $N_Y$ is increased by one for the associated batch of simulations and $K$ indexes the column including a portion of connection that is the most distant from the central column $K_S$. b) The connectivity test is negative; $N_Y$ is not increased and the efficiency metric cannot be computed. }
\label{fig:env_setting}
\end{figure}

The typical elements of a single in-silico experiment are shown in Fig. \ref{fig:env_setting}. The superimposed grid is only used to test connections among agent-populated squares and their efficiency, as explained above, and has no effect on the simulations. For each given setting, we run a batch of 100 simulations, so as to collect sufficient statistics.

\section{Results}

\begin{figure}[h!]
\centering
\includegraphics[width=\linewidth]{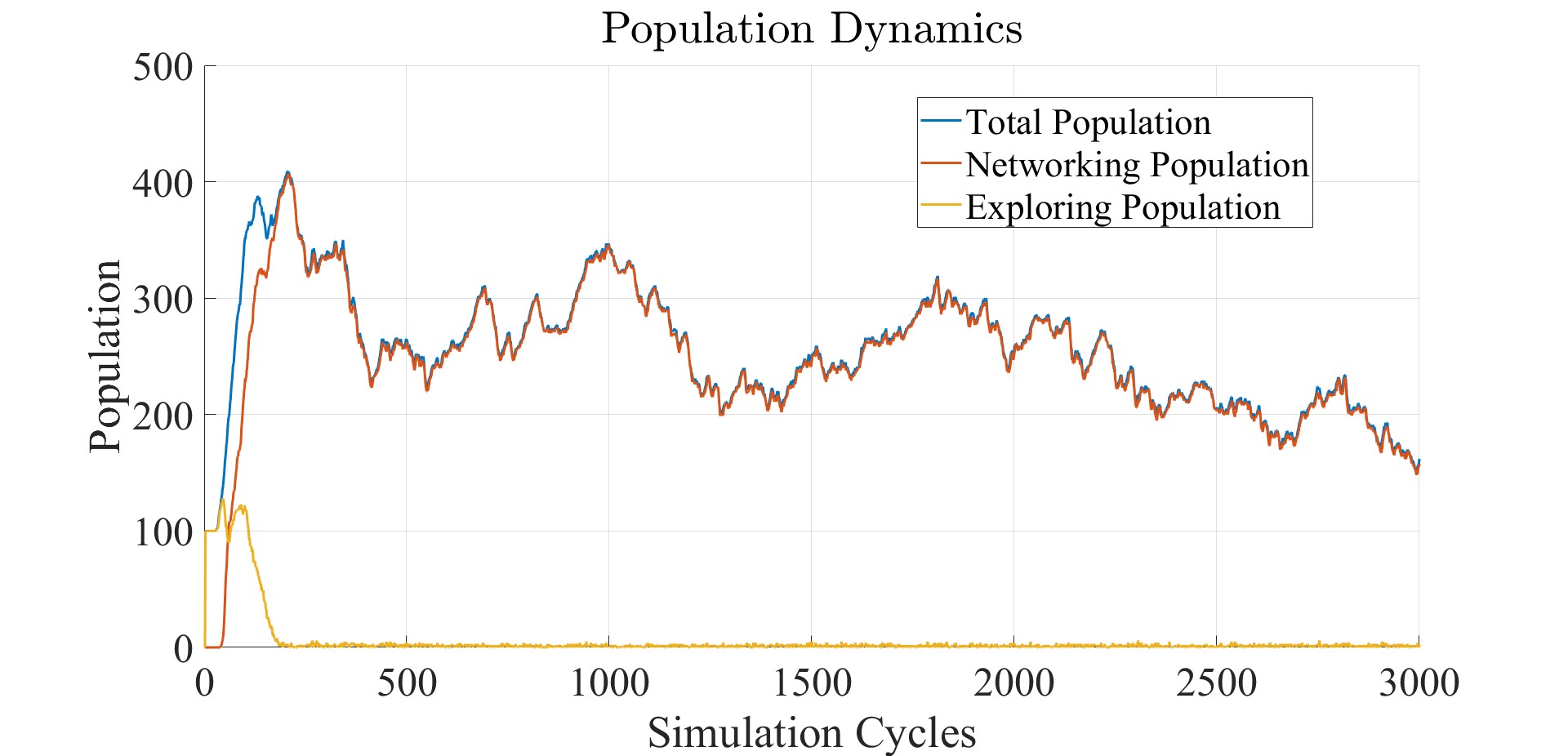}
\caption{\small Example of population dynamics for LCM. The initial exploring population mostly switches into the networking state and ends up connecting the food sources. The population initially grows and then decreases as the structure of the network stabilizes; persistent fluctuations in the total number of agents are due to demographic rules and stochasticity.}
\label{fig:curves}
\end{figure}

We compare the behavior of the models NM (which only involves networking), EM (which involves both exploration and networking, with agents able to switch between the two states) and LCM (which includes, in addition to the features of EM, demographic dynamics, as shown in Fig. \ref{fig:curves}). A simulation video for each model is reported at \url{http://giuliagiordano.dii.unitn.it/docs/papers/video_physarum.zip}.

\begin{figure*}[h!]
\centering
\includegraphics[width=\linewidth]{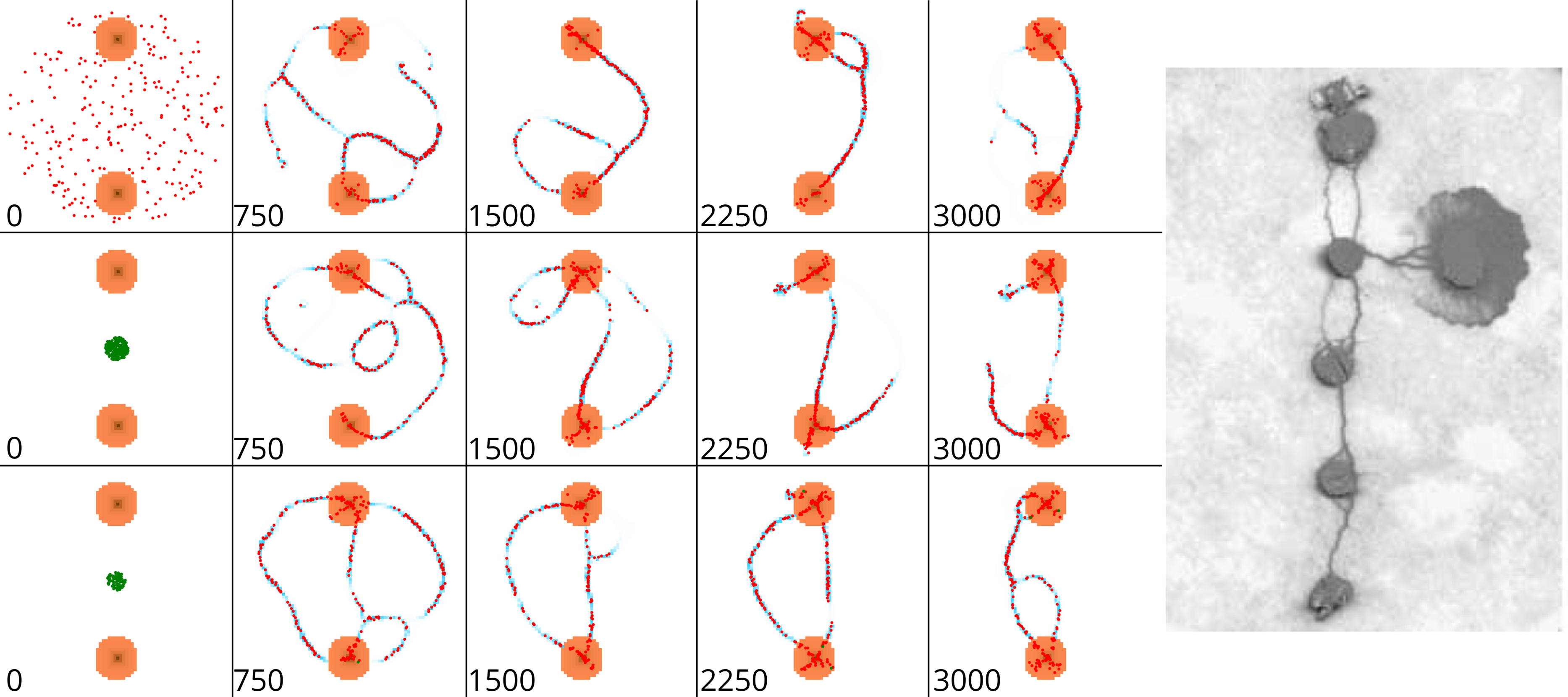}
\caption{\small (Left) Simulation snapshots for the three models (NM, top; EM, middle; LCM, bottom), using default parameters, taken at times $t=0$, $t=750$, $t=1500$, $t=2250$ and $t=3000=T$. Food sources are in orange, networking agents in red, exploring agents in green. Cells with a nonzero amount of trail tokens are in light blue. (Right) Protoplasmic tube network (vertical connections between oat flakes), plus an active plasmodium growth front forming the horizontal connection (Images courtesy of prof. Andrew Adamatzky). The tubules between pairs of flakes are reproduced by the simulations on the left.}
\label{fig:sims}
\end{figure*}

As shown in the snapshots of Fig. \ref{fig:sims}, all the models enable self-organization of the organism towards a transport network that connects food sources, even in the absence of an initial exploration phase guided by chemical gradients. Simulation videos that show the entire network formation process over time in the three cases are reported at \url{http://giuliagiordano.dii.unitn.it/docs/papers/video_physarum.zip}. For NM (Fig. \ref{fig:sims}, top), networking agents are scattered in the environment and immediately start forming connections. After an initial refinement, which removes some branches, a mature connection among food sources is formed (although a small number of agents may still fluctuate in the environment, moving away from the mature network that has formed, due to the randomness embedded in the model). EM agents (Fig. \ref{fig:sims}, middle) are initially in their exploration phase and start from a central location in the environment. However, the exploration phase is short: already after 750 cycles, they have started to form a network and rapidly connect food sources. Some disconnected agents, not contributing to the connection between the food sources, may keep existing in the long term, but they do not affect the connectedness check. LCM (Fig. \ref{fig:sims}, bottom) also involves demographic dynamics: a smaller number of initial agents can grow to quickly reproduce patterns akin to those of EM.

\begin{figure*}[ht!]
\centering
\includegraphics[width=\linewidth]{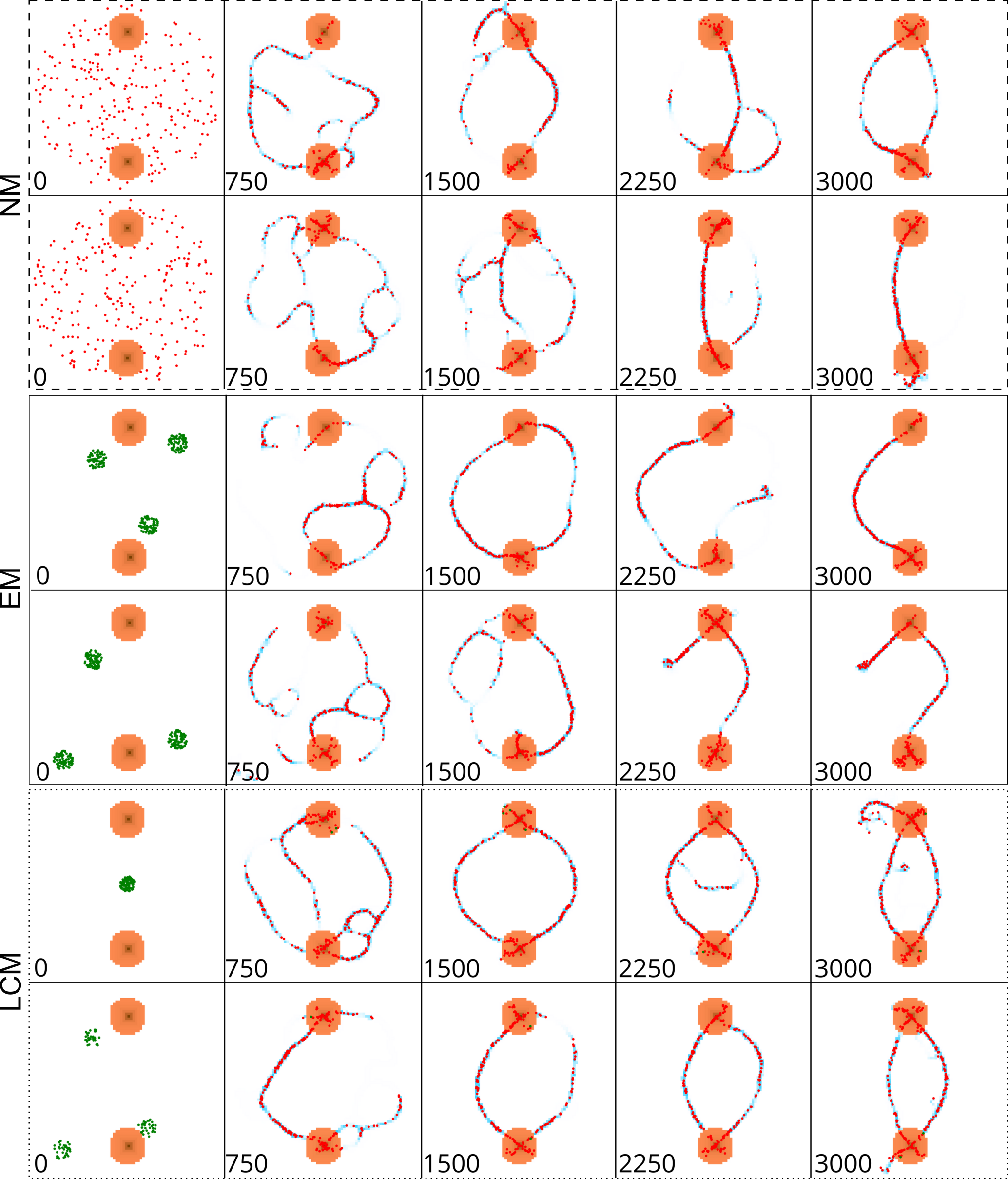}
\caption{\small Additional simulation snapshots for the three models (NM, top, in dashed frame; EM, middle, in solid frame; LCM, bottom, in dotted frame) taken at times $t=0$, $t=750$, $t=1500$, $t=2250$ and $t=3000=T$. Note that the initial configuration of the agents differs from the central location adopted in the other simulations.}
\label{fig:sims2}
\end{figure*}

Qualitatively, the mature networks formed by the three models are akin to those observed in the experiments, connecting two discrete food sources at a time; see e.g. \cite[Fig. 1]{Jones2010} and the experimental photography in Fig. \ref{fig:sims}, right. Even multi-branched connections like the one shown in Fig. \ref{fig:sims} for LCM can be observed experimentally.
It is worth stressing that our simulations include randomness and each run leads to a different outcome, including the formation of different network topologies such as those shown in Fig. \ref{fig:sims2}, which include the formation of single or double strains (similar to empirically observed ones, \textit{cf.} Fig. \ref{fig:sims}, right) or the formation of short protuberances. Also, note that Fig. \ref{fig:sims2} features simulations with initial configurations of agents that differ from the central location employed above: the initial condition only influences the early phase of the transient, then the colony converges to the same mature network. Hence, default initial conditions (as described in the Methods section) will be employed in the subsequent analysis.

\subsection{Success of connections}
\begin{figure*}[ht!]
\centering
\includegraphics[width=\linewidth]{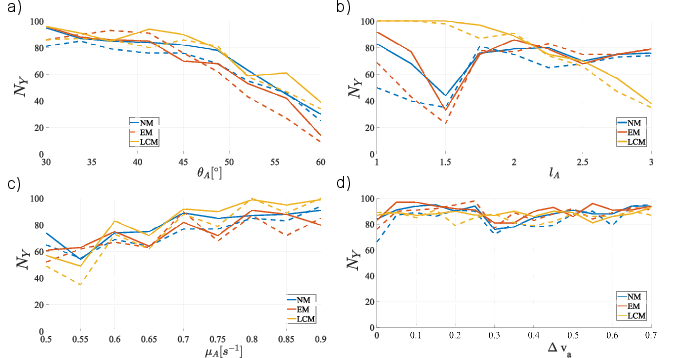}
\caption{\small Number $N_Y$ of simulations having successful connections, over 100 repeated runs, for each model (see legend) and for each parameter in its sensitivity interval (\textit{cf.} Tab. \ref{tab:default}): a) sensor arm angle $\theta_A$; b) sensor arm length $l_A$; c) trail degradation rate $\mu_A$; d) amplitude of uncertainty interval for agent speed $\Delta v_A$. Solid lines connect the values obtained using matrix $M_1$ with $n_1 = 11$, dashed lines the values obtained using matrix $M_2$ with $n_2 = 13$.}
\label{fig:success}
\end{figure*}

For a quantitative assessment of connectedness, Fig. \ref{fig:success} shows that, with default parameters, all the models yield a remarkable success rate, reaching persistent connections in a large number $N_Y$ of cases, over 100 repeated simulations.

The success rate, however, depends on parameter values. 

Across all three models, small sensor arm angles $\theta_A$ enable convergence to persistent connections almost always, while increasing $\theta_A$ significantly reduces the success rate (Fig. \ref{fig:success}a). In fact, deviations from the forward path are more likely if the agents sense trails on a wider cone, which increases the chances of connection failure.

The sensitivity of $N_Y$ with respect to $l_A$ for LCM drastically differs from that for NM and EM, as shown in Fig. \ref{fig:success}b; this difference is due to the definition of the critical density $T_N$ affecting the life cycle of agents, which depends on the sensor arm length $l_A$ (see Section \ref{sec:agents}). Short sensors enable a balanced demographic growth, while longer ones introduce demographic fluctuations that may lead to a decrease in connectivity. Since $l_A$ represents the range of influence from the surrounding environment received by a single unit of cytoplasm, our results suggest that congestion of tubular structures can be a limiting factor for the network connectivity. For NM and EM, in turn, $N_Y$ is rather constant across different values of $l_A$, except for $l_A \in [1.25; 1.75]$, where $N_Y$ plummets. In these models, demographic rules are not implemented; we thus hypothesize that having a range of interactions in this interval yields congestion of the network fluxes, not compensated by elimination of dense tracts, which in turn leads to pruning of entire veins and thus reduced connectedness. It would be interesting to confirm this hypothesis with experiments that measure the flux of nutrients passing through veins. Moreover, this seems to suggest a particular fragility of the system with respect to this spatial scale; hence, the value of $l_A$ should be carefully selected and optimized, depending on the model, in order to ensure connectedness among food sources, differently from what suggested in previous works \citep{liu2017new}.

The sensitivity to the trail evaporation rate $\mu_A$ is consistent across models: $N_Y$ tends to increase with $\mu_A$, although with small fluctuations, as shown in Fig. \ref{fig:success}c. More persistent trails sustain connections for longer and thus make the network more persistent too. The increasing trend reduces after $\mu_A = 0.8$, indicating a potential saturation point above which higher $\mu_A$ yield similar results. We notice that the default value $\mu_A = 0.7$ s$^{-1}$, taken from previous literature studies, is the value at which precisely next to the maximum of $N_Y$ reaches its maximum, thus suggesting fine-tuning.

Finally, Fig. \ref{fig:success}d shows the robustness of the self-organizing dynamics to uncertainties in the agents' speed. Apart from small fluctuations, even if individual agents have \textit{different} speed values drawn from uncertainty distributions, the organism as a whole can still achieve persistent connections. This attests to the remarkable robustness of bottom-up systems, which are able to thrive and perform complex tasks despite natural uncertainties and heterogeneity at the individual level, thanks to collective self-corrective mechanisms.

The results of the sensitivity and robustness analysis are very similar regardless of the size of matrix $M$. The observed small variations can be explained by the fact that $M_2 \in \mathbb{R}^{13 \times 13}$ has a greater sensitivity than $M_1 \in \mathbb{R}^{11 \times 11}$ in identifying empty cells.

\subsection{Efficiency of connections}

\begin{figure*}[ht!]
\centering
\includegraphics[width=\linewidth]{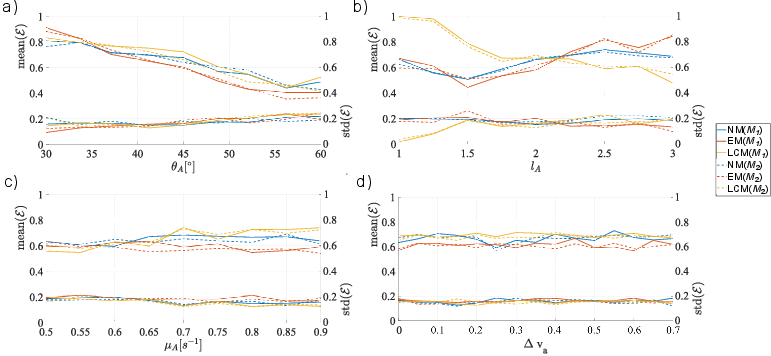}
\caption{\small Sensitivity of the normalized efficiency metric $\mathcal{E}$ in Eq. \ref{eq:efficiency_norm} (mean, top, and std, bottom of each panel, over 100 simulations) for each model (see legend) and for each parameter in its sensitivity interval (\textit{cf.} Tab. \ref{tab:default}): a) sensor arm angle $\theta_A$; b) sensor arm length $l_A$; c) trail degradation rate $\mu_A$; d) amplitude of uncertainty interval for agent speed $\Delta v_A$. Solid lines connect the values obtained using matrix $M_1$ with $n_1 = 11$, dashed lines the values obtained using matrix $M_2$ with $n_2 = 13$.}
\label{fig:measures}
\end{figure*}

The metric $\mathcal{E}$ defined in Eq. \ref{eq:efficiency_norm} allows us to quantify the efficiency of the formed interconnection, measuring its distance from a direct connection between the two food sources (vertical straight line). Fig. \ref{fig:measures} shows the mean and standard deviation (std) of $\mathcal{E}$, over 100 simulations, for each parameter set. Default parameters lead to the formation of networks with a good efficiency.

The system is robust to uncertainties $\Delta v_A$ in the agents' speed (Fig. \ref{fig:measures}d): both mean and std of $\mathcal{E}$ are approximately constant over the whole $\Delta v_A$ set. Again, the system shows a remarkable robustness to uncertainty and heterogeneity in the speed of the agents, thanks to collective mechanisms for self-correction (for instance, the density of agents adapts to spontaneously reduce speed, as it happens for traffic flows in transportation networks).

However, the outcome changes when considering sensitivity with respect to other parameters.
For all models, when increasing $\theta_A$, the network becomes more efficient (lower $\mathcal{E}$) at the price of a higher standard deviation, as shown in Fig. \ref{fig:measures}a. The comparison with Fig. \ref{fig:success}a highlights the trade-off between variability, which allows space exploration in search for more efficient solutions, and persistence, which ensures that a solution exists, albeit less efficient. Trade-offs between flexibility and guaranteed persistence of a desired behavior are typical of complex microbiological organisms \citep{proverbio2024chemotaxis}, which need to self-organize around sweet spots that compromise between two competing issues: the formation of consistent patterns, and the need to move away from local minima in search for better solutions.
A similar observation emerges from Fig. \ref{fig:measures}c, where the mean $\mathcal{E}$ slightly increases when increasing $\mu_A$ for NM and LCM, while the std remains approximately constant: increasing $\mu_A$ leads to an almost negligible loss of efficiency, while promoting the formation of more persistent connections (as shown in Fig. \ref{fig:success}c).
Finally, the sensitivity with respect to $l_A$, in Fig. \ref{fig:measures}b, shows significant differences between LCM on the one hand and NM and EM on the other hand, as it happens for $N_Y$, due to the dependence of demographic rules on the sensor arm length. For LCM, $\mathcal{E}$ decreases (hence improves) when $l_A$ increases, while the number of persistent connections decreases as shown in Fig. \ref{fig:success}b: this is another instance of biological trade-off. Conversely, for NM and EM, $\mathcal{E}$ is smaller when $l_A$ is small, and larger $\mathcal{E}$ when $l_A$ is large, suggesting a trade-off between increasing the flux through veins and preventing extremely high densities that may lead to clots.

Again, except for small fluctuations, the results are essentially independent of the size ($n_1$ or $n_2$) of matrix $M$, thus showing the consistency of our approach.

\section{Conclusion}

Self-organization is a common trait of microbiological colonies, which leverage individual basic capabilities to perform complex tasks at the macroscopic level. Using a minimal bottom-up approach, we have shown that the foraging and network formation behaviors of \textit{Physarum polycephalum} emerge from simple individual rules, and are robust against uncertainties in programs and parameters. Our findings contribute to a better understanding of such impressive feats by seemingly simple organisms, and support the development of robust bio-inspired algorithms for decentralized problem solving.

In particular, we have considered a hierarchy of bottom-up models for the foraging behavior of \textit{P. polycephalum}, and shown their remarkable robustness in yielding the emergence of transport networks. 
The simplest model, NM, is sufficient to reproduce the networking behavior and form connections among food sources with little sensitivity and high robustness, but cannot perform space exploration. Conversely, EM and LCM successfully reproduce both exploring and networking behaviors, mimicking the full foraging behavior of a self-organized organism. Their sensitivity and robustness patterns are very similar, albeit LCM more closely reproduces the actual biological mechanisms \cite{liu2017new}; however, it introduces fragility for large $l_A$ and slightly reduces efficiency.
This observation is in line with the observed trade-off between the robustness of connections and the flexibility needed to increase efficiency, already noticed in the experimental studies \citep{Nakagaki2007, Adamatzky2016}, where the organism had to self-adapt to uncertainties and obstacles. Our study paves the way to a quantitative understanding of these mechanisms, and of the trade-offs in the emergence of pattern formation.

In addition, our incremental study of complex mechanisms sheds light on the tradeoff between computational efficiency and fidelity to biological traits, which can inform bio-inspired algorithms for problem solving. Historically, bottom-up models for foraging took inspiration by other biological systems, such as the famous Ant model in \cite{dorigo2000ant}. Our approach resembles it in the way agents reinforce and follow trails of tokens. Our inclusion of autocrine chemotaxis in the trail-following scheme also resembles decentralized gathering algorithms inspired by social amoebae, see e.g. \cite{proverbio2020assessing}.

However, key differences include the development of an exploration phase, which in our model is not random but guided by chemotaxis (while it is absent in decentralized gathering schemes), and the absence of nests or centroids to return to (which are present in decentralized gathering schemes), whereas networks are spontaneously formed among the food sources. In addition, the switching between exploration and efficient networking in EM and LCM includes a further level of complexity, as it combines \textit{two} self-organized behaviors, while amoebae only exhibit spontaneous contraction towards a single spot. Our proposed models and their rich dynamics thus provide a novel and flexible complement to existing self-organized approaches for foraging and network formation on nutrient-poor substrates.
Moreover, our differently complex models have implications for technological challenges. LCM is closer to biological fidelity, and can inform future microbiological studies of population dynamics in nutrient-poor substrates, or even ecological studies about adapting behaviors in nutrient-poor environments \citep{duncan2024population, binzer2012dynamics}. However, LCM requires additional sensing capabilities and dynamic populations, which may be unfeasible when translating algorithms to artificial systems such as robot swarms \citep{markdahl2020robust, markdahl2021almost}, which rely on a fixed number of agents, and require lower complexity. Moreover, a full simulation of LCM runs in $(325 \pm 8)$s, while NM and EM, which have the same fixed number of agents, take less time to achieve a mature network:  ($301 \pm 7$)s for NM and ($300\pm8$)s for EM. This computational advantage, which becomes significant when multiple simulations are involved, can suggest the adoption of one modeling approach over the other for technological applications. In fact, the simple and local behavioral rules of EM, without any demographic requirements, are sufficient to guarantee the formation of efficient connections, with comparable or even better performance, once the parameters have been adequately chosen. EM can thus inspire lightweight algorithms to optimize traffic flows in man-made networks \cite{tokyoRailway}, test re-routing of existing solutions \cite{adamatzky2010road}, or even inspire the development of robotic swarms to carry resources on self-organized transport networks.

\begin{figure*}[ht!]
\centering
\includegraphics[width=\linewidth]{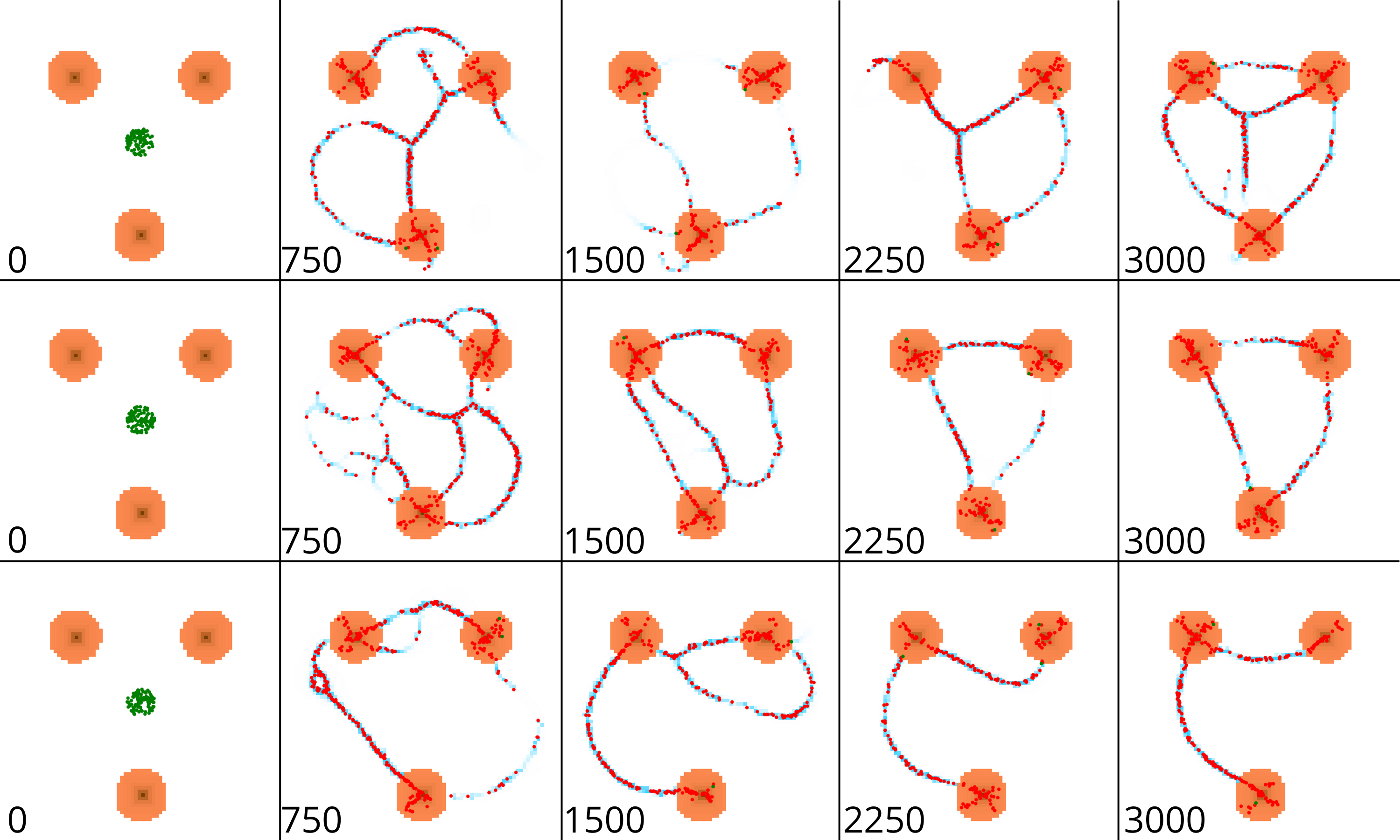}
\caption{\small Simulation snapshots of LCM experiments with a configuration with three food sources, taken at times $t=0$, $t=750$, $t=1500$, $t=2250$ and $t=3000=T$.}
\label{fig:3_food}
\end{figure*}

The present work is focused on homogeneous environments with a low number of food sources, to enable a quantitative investigation of network formation and efficiency. However, our simulation protocol is sufficiently flexible to allow the investigation of network formation among multiple food sources.  
Qualitative investigations, beyond our main scope of studying linear network formation and quantifying its efficiency, are shown in Fig. \ref{fig:3_food}, where LCM reproduces transport networks among three food sources that are close to experimental figures such as \cite[Fig. 2 and 3]{Nakagaki2004}.

Moreover, our model assumes some simplifications of \textit{Physarum}'s structure, such as the introduction of sensor arm length and angle that, albeit justified in light of abstract functions and in an application-oriented framework, have no direct biological counterparts. In addition, the model does not implement some behaviors that are known in the biology of \textit{Physarum}, but are apparently not necessary for network formation, such as self-avoidance after releasing repellents \citep{reid2012slime}. Future studies focusing on biological fidelity may refine the models by introducing these aspects and study their role in emergent dynamics, \textit{e.g.}, in improving robustness or efficiency of environment exploration.
We also stress that, in our models, reaching a food source does not imply any state switch: even if they are located at a food source, agents in the state $\sigma_1$ will produce \textit{exploring} offspring, which can possibly reach more distant food sources. This mimics the actual behavior of \textit{Physarum}, which does exhibit fully separate exploration and networking phases, and maintains an expansion front while refining a pre-formed network.

Finally, \textit{P. polycephalum} is famous for its ability to solve mazes by connecting food sources placed at their entry and exit points. Further quantifying the efficiency of such connections would improve our understanding of emergent problem solving according to \textit{optimality} criteria, and help us characterize the robustness of such algorithms in the presence of uncertainties and perturbations.

\section*{Authors' contributions}
D.R.: conceptualization, formal analysis, investigation, methodology, software, validation, visualization, writing
—original draft, writing—review and editing. 
D.P.: conceptualization, formal analysis, investigation, methodology, software, validation, visualization, writing
—original draft, writing—review and editing.
G.G.: conceptualization, formal analysis, funding acquisition, methodology, supervision, validation, writing
—original draft, writing—review and editing.

\section*{Code accessibility}
The code for MAS simulations, written in GAMA \citep{grignard2013gama}, is freely accessible at \url{https://doi.org/10.5281/zenodo.14508253}. The statistical analysis was performed with Matlab open libraries.

\section*{Ethics} This work did not require ethical approval from a human subject or animal welfare committee.

\section*{Declaration of AI use} We have not used AI-assisted technologies in creating this article.

\section*{Conflict of interest declaration.}
The authors declare no competing interests.

\section*{Funding}
This work was funded by the European Union through the ERC INSPIRE grant (project number 101076926). Views and opinions expressed are however those of the authors only and do not necessarily reflect those of the European Union or the European Research Council Executive Agency. Neither the European Union nor the European Research Council Executive Agency can be held responsible for them.

\end{document}